\documentclass[peerreview,twoside,web]{ieeecolor}
\usepackage{tmi}
\usepackage{cite}
\pdfoutput=1
\usepackage{amsmath,amssymb,amsfonts}
\usepackage{algorithmic}

\usepackage{graphicx}

\usepackage{bmpsize}
\usepackage{subcaption}
\usepackage{textcomp}
\usepackage{soul}

\usepackage{xcolor}

\def\BibTeX{{\rm B\kern-.05em{\sc i\kern-.025em b}\kern-.08em
    T\kern-.1667em\lower.7ex\hbox{E}\kern-.125emX}}
\begin{document}
%\title{Scatter-corrected restoration for spectral x-ray imaging with an energy-resolving photon-counting detector}
\title{Energy-sensitive scatter estimation and correction for spectral x-ray imaging with photon-counting detectors}
\author{Cale E. Lewis, and Mini Das*\ 
\\Department of Physics, University of Houston
 \thanks{Submitted on XX March 26, 2022. This work was partially supported by funding from the National Institute of Health/NIBIB award EB029761, the  US Department of Defense (DOD) Congressionally Directed Medical Research Program (CDMRP) Breakthrough Award BC151607, and the National Science Foundation CAREER award 1652892} \thanks{C. E. Lewis and M. Das are with the Department of Physics, University of Houston, Houston, TX 77204 USA (email: mdas@uh.edu)}}

\maketitle

\begin{abstract}
As photon counting detectors are being explored for medical and industrial imaging applications, there is a critical need to understand spectral characteristics of scattered x-ray photons. Scattered radiation is detrimental to x-ray imaging by reducing image quality and quantitative accuracy. While various scatter correction techniques have been proposed for x-ray imaging with conventional energy-integrating detectors, additional efforts are required to develop approaches for spectral x-ray imaging with energy-resolving PCDs. We show the benefits of accurate scatter estimation and correction for each energy bin when using a photon counting detector. We propose a scatter estimation model that accounts for the energy-dependent scatter characteristics in projection imaging. This can then be used to restore quantitative accuracy for spectral x-ray imaging with PCDs. Results are shown in the context of contrast-enhanced spectral mammography using dual-energy subtraction to digitally isolate iodine targets (2.5--40 mg/ml). In the presence of scatter, the projected iodine densities are increasingly underestimated as the object thickness increases. The energy-sensitive scatter correction improves the iodine density estimation up to 46\%. These results suggest that our scatter estimation model can accurately account for the energy-dependent scatter distribution, which can be an effective tool for scatter compensation in spectral x-ray imaging. Implementing this scatter estimation model does not require any modifications to the acquisition parameters and is transferable to other x-ray imaging applications such as tomosynthesis and CT.
\end{abstract}


\section{Introduction}
\label{sec:introduction}

\IEEEPARstart{P}{hoton} counting detectors (PCDs) with energy-resolving capabilities show promising advantages for x-ray imaging \cite{taguchi2013vision, shikhaliev2008computed, schmidt2009optimal}. Spectroscopic x-ray imaging with PCDs can be applied for quantitative material identification and characterization such as K-edge imaging \cite{roessl2007k, torrico2019single}, material decomposition \cite{lee2014quantitative, fredette2019experimental, fredette2019multi}, and phase-contrast imaging \cite{gursoy2013single, das2014spectral, vazquez2020quantitative}. However, scattered radiation is a well-known detriment to x-ray imaging resulting in increased image noise and reduced quantitative accuracy \cite{bushberg2011essential, siewerdsen2001cone}. Additionally, incoherently scattered radiation undergoes energy transfer with the medium and therefore obscures the spectral characteristics of the detected primary radiation. Previous investigations have demonstrated the detrimental impact of scattered radiation in spectroscopic imaging with PCDs \cite{wiegert2009impact, schmidt2010ct, lewis2017impact, lewis2019spectral, sossin2014influence, sossin2018experimental}.

Developing scatter suppression techniques to account for these detriments is a long-standing problem which has many proposed approaches \cite{ruhrnschopf2011general}. Pre-processing techniques employ additional hardware components to prevent the scatter radiation from reaching the detector. Collimating the incident beam reduces the scatter-to-primary ratio (SPR) \cite{boone2000scatter} but at the cost of a smaller field-of-view, which may be impractical for some applications. More sophisticated collimation techniques such as slot-scanning \cite{jing1998scattered} and multiple-slit \cite{aslund2004scatter} systems have been developed to overcome these restrictions. However, these systems tend to increase acquisition time. This may also result in motion artifacts and imposing stringent alignment and mechanical precision requirements. Increasing the air-gap (object-to-detector distance) provides a simple and practical approach to reduce the SPR \cite{boone2000scatter}, but may be limited by spatial restriction of the imaging system and geometric blurring of the finite focal spot of the x-ray source. Anti-scatter grids (ASGs) have shown to effectively reduce scatter in radiography \cite{kalender1982calculation, chan1990studies} and dedicated breast CT \cite{kwan2005evaluation}. However, there is partial absorption of the primary intensity \cite{boone2002grid}. The increased quantum noise may undermine the reduced scatter and therefore make ASGs impractical, particularly for high-resolution small-pixel detectors \cite{siewerdsen2004influence, singh2014limitations}. 

On the other hand, post-processing techniques attempt to recover the primary radiation within the total measured intensity through a scatter estimation and compensation technique. These estimation techniques may be categorized into measurement-based methods and model-based methods. Measurement-based methods include sampling the scattered radiation during image acquisition by suitable modifications in the beam path. As an example, the collimator-shadow technique \cite{siewerdsen2006simple} consists of estimating the scatter within the attenuated region of a collimator placed in front of the sample and interpolating the scatter in the unattenuated region. However, this interpolation is inaccurate for objects with strong asymmetry or heterogeneity. Similarly, beam-stop arrays sample the scatter distribution behind a fixed attenuator grid placed in front of the sample. However, this approach often requires two projections (with and without the grid) to accurately obtain the spatial scatter distribution from object. In the context of spectral x-ray imaging with PCDs, the use of partially-attenuating grids \cite{sossin2016novel, sossin2016experimental} and primary modulators \cite{pivot2020scatter} have been proposed. However, introducing these additional components to the beam path often alters the patient dose and also results in image artifacts. 

Model-based scatter estimation techniques rely on analytic models or Monte Carlo simulation. However, pure analytical or simulation methods are computationally expensive and require detailed \textit{a priori} knowledge of the imaging system (including the x-ray source, detector and object). Kernel-based methods attempt to reduce these demands through clever approximations and form tractable analytic solutions. Beam-scatter-kernel methods approximate the magnitude and angular spread of the scattered radiation produced by a pencil-beam through the object \cite{barrett1996radiological, boone1986characterization, kruger1994regional, sun2010improved}. While kernel-based scatter estimation methods have shown promising improvements when using conventional energy-integrating detectors (EIDs), these methods have not yet been investigated for spectroscopic x-ray imaging with PCDs.

We have previously characterized the spectral behavior of scattered radiation for energy-resolved x-ray imaging with PCDs \cite{lewis2019spectral}. Here, we present methods to apply these spectral scatter signatures to develop an energy-sensitive scatter estimation technique using a kernel-based approach. This estimation is then used for scatter correction in energy-resolved x-ray images. The benefits of this new approach is presented in the context of contrast-enhanced spectral mammography using dual-energy subtraction (DES) to digitally isolate iodine targets. First, we report the impact of the scattered radiation in the quantitative estimation of the iodine targets. Next, we demonstrate quantitative restoration of the iodine targets using the proposed energy-sensitive scatter estimation and compensation. 

\section{MATERIALS AND METHODS}

This study is concerned with the impact of scattered radiation in the context of spectral x-ray imaging using dual-energy subtraction. The mathematical formalism of dual-energy subtraction and the expected impact of scattered radiation are presented in the Appendix. Section \ref{sec:theory} describes an energy-sensitive scatter estimation model, which is then used for scatter correction. Section \ref{sec:exp} details the experimental procedure used for the data acquisition and estimation of the reference scatter fractions.

\subsection{Scatter estimation model}\label{sec:theory}

An energy-dependent estimate of the scattered radiation $\tilde{I}_s(E)$ can be used for scatter correction of the  measured intensity $I(E)$ and obtain a primary radiation estimate:
	\begin{equation}
		\tilde{I}_p(\pmb{x},E) = I(\pmb{x},E) - \tilde{I}_s(\pmb{x},E),
	\end{equation}
	where $\pmb{x}=(x,y)$ are the coordinates at the detector plane and the tilde corresponds to an estimated quantity. The scattered radiation is composed of coherently and incoherently scattered x-rays. The coherently scattered radiation is typically restricted to small-angle deflections that occur predominately at low energies ($<\sim$20 keV for low-Z materials). The incoherently scattered radiation plays a proportionally larger role than the coherent scatter, and is characterized by a relatively broad angular distribution and an energy transfer with the medium. The spectral properties of the incoherent scatter radiation are typically overlooked for conventional EID x-ray imaging, where the recorded signal corresponds to the mean energy characteristics of the total collected x-ray photons. On the other hand, x-ray imaging with energy-resolving PCDs benefits by accounting for the spectrally-varying scatter properties and permitting more precise scatter estimation and compensation.
	
\begin{figure*}
        \centering
        \begin{subfigure}[]{0.40\textwidth}
            \centering
             \includegraphics[width=\textwidth]{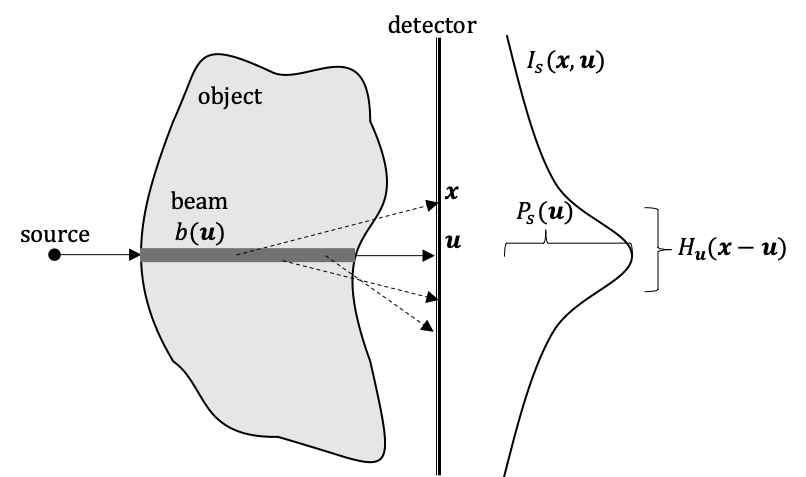}
            \caption{}
            \label{fig:bsk}
        \end{subfigure}
        \begin{subfigure}[]{0.3\textwidth}
            \centering
             \includegraphics[width=\textwidth]{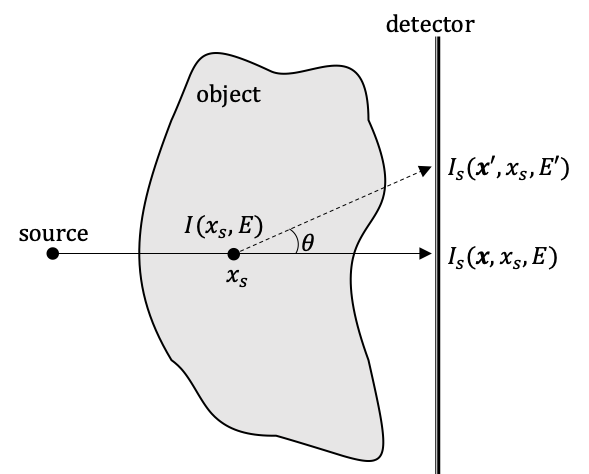}
            \caption{}
            \label{fig:first_order}
        \end{subfigure}
        \caption{(a) Schematic of the beam-scatter-kernel scatter model. (b) Schematic for first-order scatter estimation model resulting from object element $x_s$.}
	\end{figure*}

	An energy-sensitive scatter estimation is derived using the beam-scatter-kernel approach. Using this approach, a line element (beam) through the object along a primary axis $\pmb{u}=(u,v)$ produces a scatter profile $I_s(\pmb{x}, \pmb{u})$ across the detector, as shown in Figure \ref{fig:bsk}. Each individual scatter profile can be modeled to have a scatter potential $P_s$, relating to the magnitude of scatter generated along the beam that reaches the detector, and a blurring kernel $H_{\pmb{u}}$, relating to the angular spread of the scatter generated along the beam. The full scatter distribution can be obtained as the superposition of the individual scatter profiles produced by all of the beams through the object:

	\begin{equation}
	 \tilde{I}_s(\pmb{x},E)= \iint_{\pmb{u}} P_s(\pmb{u},E) \times H_{{\pmb{u}}}(\pmb{x}-\pmb{u}, E) d\pmb{u}.
	\end{equation}	
	
Generally, the spatial variations of the scattered radiation are constrained to low frequencies ($<$$\sim$0.1 cm$^{-1}$ \cite{barrett1996radiological, bootsma2013spatial}) and occur on the order of a few centimeters. This results in a smooth scatter distribution that slowly varies across the full projection. Our investigations are limited to small region-of-interests which are less impacted by these spatial variations, and therefore we disregard the blurring kernel $H_{\pmb{u}}$. As a result, the estimated scattered radiation distribution can be simplified to depend only on the scatter potential:

	\begin{equation}\label{eq:is_approx}
		\tilde{I}_s(\pmb{x},E) \approx P_s(\pmb{x},E).
	\end{equation}	
	
	The scatter potential $P_s$ can be described by a first-order scatter estimation model \cite{bai2000slice, rinkel2007new, wiegert2005model}, which is schematically represented in Figure \ref{fig:first_order}. This model can be divided into three steps:
	
	i) The x-ray intensity reaching a position $x_s$ within the object is described by:	
		\begin{equation}
			I(x_s, E) = I_0(E) e^{-\int_0^{x_s} \mu(x, E) dx}.
		\end{equation}
	
	ii) The probability of a scatter interaction at position $x_s$ is described by the Klein-Nishina differential cross-section $\frac{d \mu_s}{d \Omega}(x_s, \theta)$. 
	
	iii) The number of scattered photons produced at $x_s$ that are deflected by an angle $\theta$ and reach the detector plane is proportional to the attenuation through the rest of the object: $e^{-\int_{x_s}^{x'} \mu(x, E') dx}$, where $E'(\theta, E)= \frac{E}{1+\frac{E}{m_0}(1- cos \theta)}$ and $m_0$ is the rest mass of an electron. 
	
	Combining these three steps complete the estimation of the first-order scatter $I_{s1}$ generated by a point $x_s$ within the object that reaches the detector plane at an angle $\theta$:
		\begin{align}\label{eq:first_order_model}
			I_{s1}(x_s, \theta, E') &= I(x_s, E) \times \frac{d \mu_s}{d \Omega}(x_s, \theta) \times e^{-\int_{x_s}^{x'} \mu(x, E') dx}.
		\end{align}
	However, this solution is not easily computed due the dependence on the last term. A tractable solution is obtained by assuming that the transmission fraction of the scattered x-ray is approximately equal to that along the primary axis (i.e. $\theta=0$) such that:
	\begin{equation}\label{eq:tf_approx}
		e^{-\int_{x_s}^{x'} \mu(x,E') dx} \approx e^{-\int_{x_s}^{x} \mu(x,E) dx}.
	\end{equation}

	Combining the first-order scatter estimation (Equation \ref{eq:first_order_model}) and this approximation (Equation \ref{eq:tf_approx}) results in a simplified solution:
		\begin{align}
			I_{s1}(x_s, \theta, E) &\approx  I(x_s, E) \times \frac{d \mu_s}{d \Omega}(x_s, \theta) \times e^{-\int_{x_s}^{x} \mu(x, E) dx} \nonumber\\
			&= I_0(E) e^{-\int_{0}^{x} \mu(x, E) dx} \times \frac{d \mu_s}{d \Omega}(x_s, \theta) \nonumber\\
			&= I_p(\pmb{x},E) \times \frac{d \mu_s}{d \Omega}(x_s, \theta).
		\end{align}	
	This expression suggests that the amount of first-order scatter generated by the point $x_s$ into an angle $\theta$  can be approximated by the product of the primary radiation intensity reaching the detector plane (that passes through $x_s$) and the Klein-Nishina differential cross-section. 
		
	The total scatter intensity reaching the detector due to the point $x_s$ is determined by summing the scatter interactions in the forward direction (i.e. toward the detector):
		\begin{align}
			I_{s1}(x_s, E) &= I_p(\pmb{x},E) \times \int_{\theta=-\frac{\pi}{2}}^{\frac{\pi}{2}} \frac{d \mu_s}{d \Omega}(x_s, \theta) d\Omega \nonumber\\
						&= I_p(\pmb{x},E) \times \mu_s^f(x_s, E),
		\end{align}
		where $\mu_s^f(x_s, E)$ is the scattering cross-section coefficient for forward-directed deflections.

	Finally, the full first-order scatter intensity at a detector position $\pmb{x}$ is determined by summing $I_{s1}(x_s, E)$ for all points $x_s$ along the beam:
		\begin{align}
			I_{s1}(\pmb{x},E) &= \int_0^{\pmb{x}} I_{s1}(x_s, E) dx_s \nonumber\\
			&\approx I_p(\pmb{x},E)\times \int_0^t \mu_s^f(t',E)dt',
		\end{align}
		where $t$ is the object thickness through a given line element along the primary axis. This expression completes the first-order scatter distribution which can be approximated as the product of the primary radiation intensity reaching the detector plane and the `forward-directed' scattering cross-section coefficients along the beam path. However, in practice, both the primary intensity and distribution of the scattering cross-section coefficients are unknown.
		
	\begin{figure*}[h!]
		    \centering
		  	\includegraphics[width=1.0\textwidth]{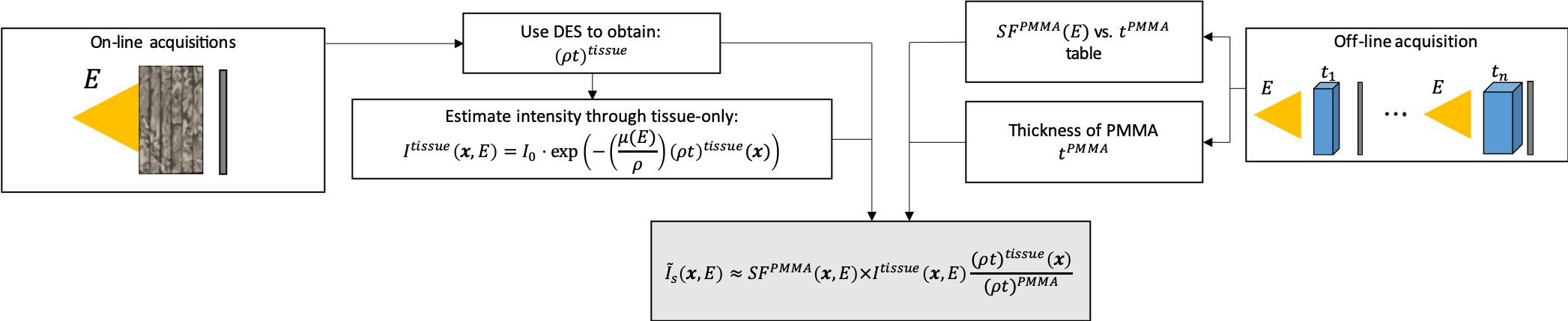}
		    \label{fig:}   
		\caption{Flowchart depiction of the scatter estimation procedure.}
		\label{fig:flowchart}
	\end{figure*}	
%	\begin{figure}[h!]
%		    \centering
%		  	\includegraphics[width=0.49\textwidth]{fig/flowchart_2}
%		    \label{fig:}   
%		\caption{Flowchart depiction of the scatter estimation procedure.}
%		\label{fig:flowchart}
%	\end{figure}		
			
	Instead, the first-order scatter through homogeneous objects can easily be approximated. Considering a homogeneous PMMA slab of thickness $t^{PMMA}$, the first-order scatter is described by: 
	\begin{equation}
	I_{s1}^{PMMA} =  I^{PMMA}_p \times (\mu_s^f)^{PMMA} \times t^{PMMA},
	\end{equation}	
	\textcolor{black}{where the energy-dependence of the intensities $I(E)$ and scattering coefficients $(\mu_s^f(E))$ are implied.} The attenuation properties of PMMA are relatively similar to other low-Z materials such as soft tissue. These similarities will be used   to treat the scatter radiation from PMMA slabs as a reference for the scatter distribution of the unknown object. This relationship can be determined by assuming that the ratio of the first-order to multiple-order interactions are approximately equal for the unknown object and PMMA slabs:
	\begin{equation}\label{eq:first_to_multi}
	\frac{\tilde{I}_{s1}}{\tilde{I}_{s}} \approx \Big(\frac{I_{s1}}{I_{s}} \Big)^{PMMA}.
	\end{equation}
	This approximation is suitable when assuming that the scatter cross-sections are similar for the PMMA and soft tissue, and that \textcolor{black}{the spatial distribution of the multiple-order scatter of each material is slowly-varying.} \textcolor{black}{For an object with contrast agent targets, it is assumed that the scattered radiation is predominately due to the soft tissue background, given that the contrast agent targets produce a negligible amount of scattered radiation due to their low-concentration and proportionally smaller object volume.}
	
	The full scatter distribution of the soft tissue object can be determined by rearranging Equation \ref{eq:first_to_multi} and substituting in the corresponding first-order scatter estimation expressions:
	\begin{equation}\label{eq:Is_general_mu}
	\tilde{I}_s \approx I_s^{PMMA} \bigg[ \frac{I_p \times \int_0^t \mu_s^f(t')dt'} {I_p^{PMMA} \times (\mu_s^f)^{PMMA} \times t^{PMMA}} \bigg].
	\end{equation}
	However, this expression relies on the primary intensity through the object and the PMMA slabs which is unknown. This can be addressed by assuming that the ratio $I_p / I_p^{PMMA} \approx I / I^{PMMA}$. This is a suitable approximation if the added scatter signal produces low spatial variations to the total measured intensity. The benefit of this expression is that the scatter produced by the soft tissue object can be approximated with the total measured intensity of the unknown object and simple measurements with homogeneous PMMA slabs. The scatter distribution of the PMMA slabs ($I_s^{PMMA}$) can be determined with a variety of methods, including measurement-based methods or Monte Carlo simulation. However, this expression also relies on an estimation of the forward-scatter cross-section distribution $\mu_s^f$ which is generally unknown.
	
	Previous literature \cite{rinkel2007new} estimates the $\mu_s^f$ distribution by an initial CT reconstructed image. However, this estimation cannot be made within projection imaging. Alternatively, the following approximations are prescribed to obtain the scatter distribution within projection imaging of low-Z materials. The object is assumed to consist primarily of soft tissue such that the first-order scatter radiation can be determined by:

	\begin{equation}\label{eq:homo_approximation}
	I^{tissue}_{s1} \approx I \times (\mu_s^f)^{tissue} \times t^{tissue}.
	\end{equation}		
	This expression suggests that the first-order scatter through the soft tissue can be approximated by the product of the total measured intensity ($I$), the known `forward-direction' scatter cross-section, and known object thickness. However, small amounts of high-Z materials (such as diluted iodine) reduce the total measured intensity which causes an underestimation of the first-order scatter and severe spatial variations. 
	
	Instead, the total measured intensity $I$ through the sample  can be replaced by an estimation of the total intensity through the low-Z material only:
	\begin{equation}\label{eq:i_tiss}
		I^{tissue}(\pmb{x}, E) = I_0(E) \times e^{-\frac{\mu(E)}{\rho} \cdot [\rho t(\pmb{x})]^{tissue}}.
	\end{equation}
	This estimation can be obtained by an initial approximation of the projected tissue density map $(\rho t)^{tissue}$ obtained by dual-energy subtraction (see Equation \ref{eq:bg_map}) and the mass attenuation coefficient $\frac{\mu(E)}{\rho}$ obtained from the NIST XCOM database \cite{berger2010xcom}.

	Combining Equations \ref{eq:Is_general_mu}, \ref{eq:homo_approximation} and \ref{eq:i_tiss}, the scattered radiation distribution through the unknown object is described by: 
	\begin{equation}
	\tilde{I}_s \approx I_s^{PMMA} \bigg[ \frac{I^{tissue} \times \Big(\frac{\mu_s^f}{\rho}\Big)^{tissue} \times (\rho t)^{tissue}} {I^{PMMA} \times \Big(\frac{\mu_s^f}{\rho}\Big)^{PMMA} \times (\rho t)^{PMMA}} \bigg],
	\end{equation}	
	Assuming that the scattering coefficients are similar for the soft tissue and PMMA ($(\mu_s^f / \rho)^{tissue} \approx (\mu_s^f/ \rho)^{PMMA}$), the final expression for the estimated scatter distribution can be determined by:
	%A similar result is reported in \cite{dinten2004x}.
	\begin{equation}\label{eq:is}
	\tilde{I}_s(\pmb{x},E) \approx SF^{PMMA}(\pmb{x},E) \times I^{tissue}(\pmb{x},E) \frac{[\rho t(\pmb{x})]^{tissue}}{[\rho t]^{PMMA}}.
	\end{equation}	

	In review, this estimation relies on a few measurable quantities, depicted in Figure \ref{fig:flowchart}. As shown in our recent work \cite{lewis2019spectral}, the required spectrally-varying scatter fraction values of homogeneous PMMA slabs ($SF^{PMMA}(\pmb{x},E)$) can be determined through measurement-based methods or Monte Carlo simulation \cite{lewis2019spectral} using equivalent acquisition parameters as with the unknown object. The PMMA slabs are chosen with a thickness $t^{pmma}$ equivalent to the object to produce similar amount of scattered radiation. The scatter due to the anatomical bulk is isolated by considering the soft tissue density projection ($[\rho t(\pmb{x})]^{tissue}$) which can be estimated with an initial dual-energy subtraction.

%	Finally, the corrected primary intensity $\tilde{I}_p$ is obtained by subtracting the estimated scatter $\tilde{I}_s$ from the total measured intensity $I$:
%	\begin{equation}
%		\tilde{I}_p(\pmb{x},E) = I(\pmb{x},E) - \tilde{I}_s(\pmb{x},E).
%	\end{equation}
%	An accurate estimation of the scattered radiation will improve the result in the corrected primary intensity to 

\subsection{Experimental setup}\label{sec:exp}

\subsubsection{Sample acquisition} 
\begin{figure}
		\centering
		    \includegraphics[width=0.49\textwidth]{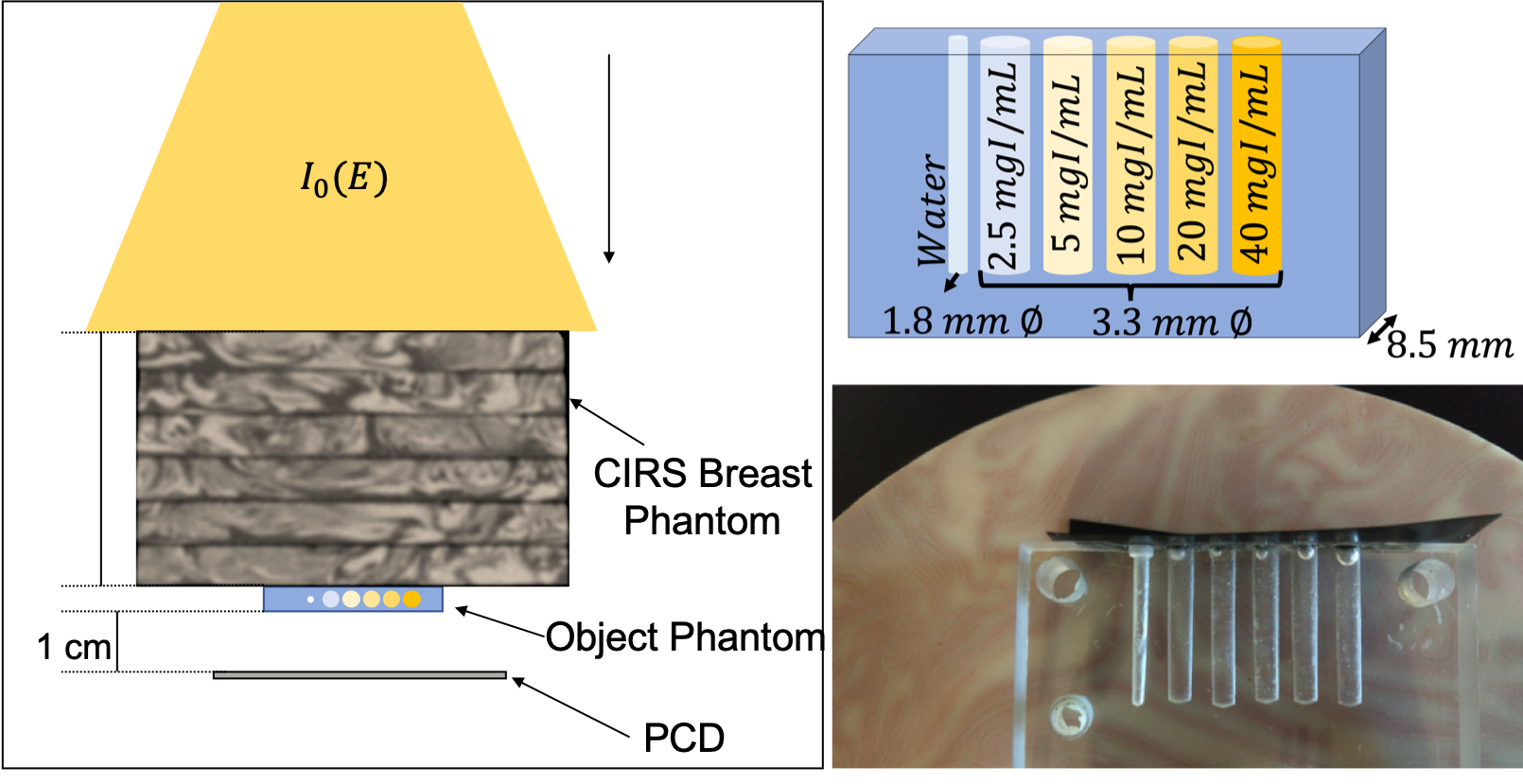}
		    \label{fig:}   
		  \caption{(left) Benchtop schematic. (upper right) Schematic of the contrast agent phantom consisting of a PMMA slab with iodine-filled cavities. (bottom right) Image of the contrast agent phantom with breast phantom background.\label{fig:bench}}
		\end{figure}

Experiments were conducted on a bench-top x-ray imaging system to obtain projections similar to those in mammography (see Figure \ref{fig:bench}). X-rays were generated using a Hamamatsu microfocus X-ray tube unit (L8122-01) consisting of a tungsten anode target with a 200 $\mu$m thick beryllium output window. The tube was operated at 60 kVp with a 500 $\mu$A current and a 50 $\mu$m focal spot. 

	The contrast agent phantom consisted of an 0.85 cm thick PMMA slab with five cylindrical cavities of 0.33 cm diameter. These cavities were filled with varying dilutions of iodine (Optiray320: 320 mg/ml) for concentrations including 2.5, 5.0, 10, 20, and 40 mg/ml. Mammographic anatomical noise was emulated with varying thicknesses of Model 020 BR3D breast imaging phantoms (CIRS, Inc., Norfolk, VA). \textcolor{black}{Acquisitions were obtained for the contrast agent phantom without additional thickness and also obtained including 2 and 5 cm thickness of additional BR3D breast imaging phantom. For achieving cases of larger object thickness (8.85cm and 10.85cm), homogeneous slabs (with thicknesses 3cm and 5 cm) of PMMA were added to the heterogeneous BR3D breast phantom (which is 5cm thick plus the 0.85cm thick contrast agent).}
	
	The detector consisted of a 5$\times$1 tiled array of Medipix3RX PCDs. All units had a 1000 $\mu$m thick CdTe substrate to convert incident radiation into electrical signal and contains a 256$\times$256 pixel array of 55 $\mu$m pixel pitch. \textcolor{black}{The total detector area measures 1.408 cm $\times$ 7.040 cm.} The detector was placed 70 cm from the source and with 1 cm spacing from the object. The detector was operated in charge-summing mode to improve the detector response from charge-sharing events. Acquisitions were obtained using energy thresholds of 19, 27, 36, and 44 keV with 1 min acquisition time per threshold. Each pixel value in these projections contain the number of incident photons with an energy deposition above the applied threshold. The number of photons in a given energy window are estimated by matrix subtraction of two adjacent thresholds. For example, the number of incident photons between 19 keV and 27 keV are estimated by subtracting the acquisition with the 19 keV threshold by the acquisition with the 27 keV thresholds. A 3$\times$3 median filter was also applied to the projections of each energy window in order to reduce the image noise.

\subsubsection{Simulated scatter fraction reference values} 
\begin{figure}[]
		    \centering
		    \includegraphics[width=0.49\textwidth]{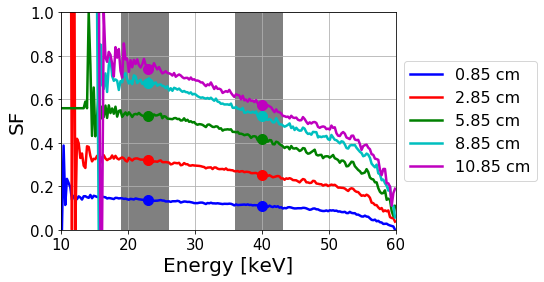}
		    \label{fig:}   
		\caption{Spectrally-resolved scatter fractions obtained from Monte Carlo simulation for different thicknesses of homogeneous PMMA slabs. The highlighted regions correspond to the energy windows used in this study. The solid circles represent the mean scatter fraction within a given energy window.}
		\label{fig:sf}
	\end{figure}

Scatter fraction ($SF$) values through homogenous PMMA slabs are required as a reference for the scatter estimation model of the unknown object (see Equation \ref{eq:is}). Monte Carlo simulation was used to precisely determine these spectrally-dependent $SF$ values. Simulations were prepared with the BEAMnrc software package \cite{kawrakow2001egsnrc} using equivalent parameters as the experimental acquisition except for replacing the sample with a homogeneous PMMA slab of equivalent area and thickness. \textcolor{black}{An ideal energy-sensitive detector is modeled in $SF$-estimation.} Generally, the spatial distributions of $SF(\pmb{x},E)$ are obtainable through Monte Carlo simulation. Rather, we assume that the $SF(\pmb{x}, E)$ does not exhibit significant spatial variations within the imaged region. Figure \ref{fig:sf} shows the spectrally-resolved scatter fraction values obtained through the simulation for different thicknesses of PMMA. 

%%%%%%%%%%%%%%%%%%%%%%%%%%%%%%%%%%%%%%%%%%%%%%%%%%%%%%%%%%%%%%%%%%%%
% RESULTS
%%%%%%%%%%%%%%%%%%%%%%%%%%%%%%%%%%%%%%%%%%%%%%%%%%%%%%%%%%%%%%%%%%%%%
\section{RESULTS}

Here, we present our results on spectral scatter correction applied to dual-energy subtraction imaging.

\subsection{Dual-energy subtraction}
\begin{figure}[]
		    \centering
		    \includegraphics[width=0.49\textwidth]{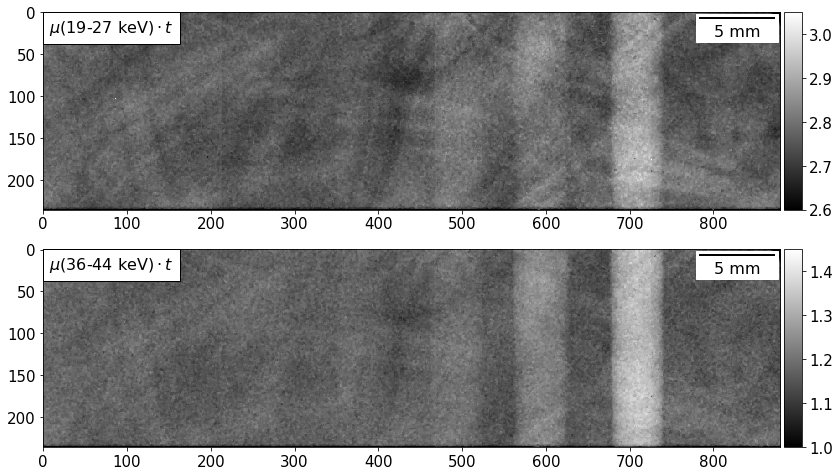}
		\caption{Logarithmic intensity projections of the phantom with additional 5 cm thick anatomical background for (top) 19--27 keV and (bottom) 36-44 keV energy windows.}
		\label{fig:mut}
		\end{figure}

Figure \ref{fig:mut} shows negative logarithmic intensity projections obtained within two energy windows of the contrast agent phantom described in Section \ref{sec:exp}. In each projection, the iodine targets become increasingly obscured by the anatomical structures as the concentration is reduced. These background structures also disrupt the perceived target rod uniformity and increase ambiguity of the iodine distributions. Increasing the energy window above the iodine K-edge (33.2 keV) enhances the target rods due to the increased photoelectric absorption (see Figure \ref{fig:lac}). However, the overlapping anatomical structures still interfere with the visibility of the target rods and concentration quantitation is uncertain.

\begin{figure}[]
		    \centering
		    \includegraphics[width=0.49\textwidth]{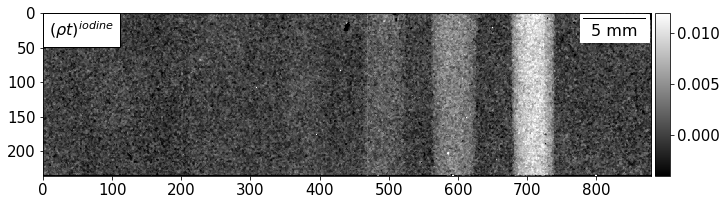}
		\caption{Projected iodine density obtained through dual-energy subtraction of the phantom with 5 cm thick anatomical noise background.}
		\label{fig:de_result}
	\end{figure}

\textcolor{black}{The results of dual-energy subtraction (Equation \ref{eq:rhoc} in the Appendix) is presented. For clarity, the steps used are reiterated. First the spectral measurements are obtained to get $\mu(E_1)$ and $\mu(E_2)$ for the two desired energy bins (see Figure \ref{fig:mut}). Next, using these as inputs, Equation \ref{eq:rhoc} is implemented to get the contrast agent density map ($[\rho t]_c$) as shown in Figure \ref{fig:de_result}}. The visibility of the iodine targets is greatly improved with the digital removal of the overlapping anatomical structures and the lower concentration target rods are more easily discernible against the uniform background. Visualizing the rod uniformity is also improved, which was previously ambiguous in the negative logarithmic intensity projections (see Figure \ref{fig:mut}).

\textcolor{black}{Figure \ref{fig:uncorrected} shows image profiles, computed as the average of the line profiles over the rod length, for the projected iodine densities using different sample thicknesses. The top plot shows the effect of unresolved spectral errors (due to the CdTe sensor in PCD) along with scatter related errors in the process. With the prior knowledge that the background regions (non-iodinated) has zero iodine density, one can implement a simple base-line correction which accounts for the spectral errors. The result of this is shown in the middle plot which effectively isolates just the effects of the scattered radiation and is described in more detail in the following section. Such baseline correction methods can be effective for such applications like contrast enhanced mammography when the experimenter is aware of a contrast free region within the region of interest.}
	
%	The anatomical background and iodine target rods can be computationally separated using dual-energy subtraction, as shown in Figure \ref{fig:de_result}. The tissue distribution corresponds to the overlaying background structures observed in the negative logarithmic intensity projections (see Figure \ref{fig:mut}). The measured quantities underestimate the expected product of the thickness and density (5.85 cm $\times$ 1.19 g$\cdot$cm$^{-1}\approx$ 7.0 g/cm$^2$) due partly to the approximations in the decomposition model and signal distortions due to scattered radiation and imperfect detector response. Nonetheless, the isolated iodine target rods can be recovered in the contrast map, which greatly improves visibility of the iodine distributions. Visibility of the rod uniformity is now improved, which was previously ambiguous in the negative logarithmic intensity projections (see Figure \ref{fig:mut}). With the background structures removed, the lower concentration target rods are more easily discernible against the uniform background.  

%		\begin{figure}[]
%		    \centering
%		    \includegraphics[width=0.45\textwidth]{fig/de_result}
%		\caption{Dual-energy subtraction performed using the projections of Figure \ref{fig:mut} separating the (top) anatomical background and (bottom) iodine.}
%		\label{fig:de_result}
%		\end{figure}	
		
\subsection{Effects of scattered radiation}
\begin{figure}[]
		    \centering
		    \includegraphics[width=0.48\textwidth]{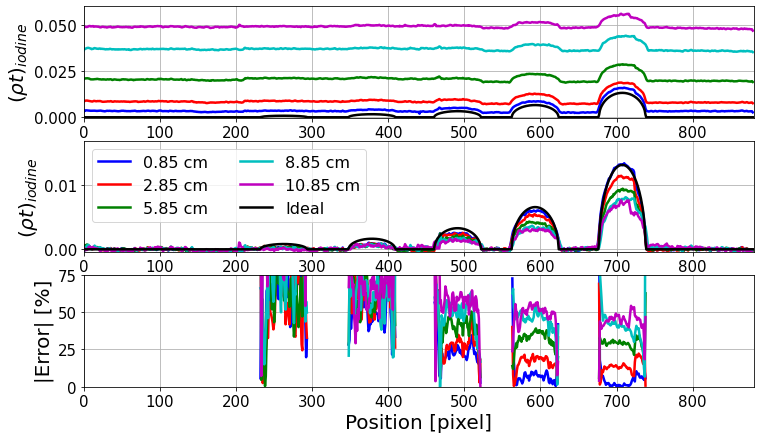}
		\caption{\textcolor{black}{(top) Estimated profiles of iodine maps before spectral errors are unaccounted for. (\st{top} middle) After prior-knowledge assisted baseline shifting has isolated errors due to scatter only and (bottom) the corresponding percent errors due to scatter only as they vary with sample thicknesses. As seen quantitative accuracy reduces with increasing sample thickness due to an increasing scatter fraction.}}
		\label{fig:uncorrected}
	\end{figure}

\textcolor{black}{The image profiles in Figure \ref{fig:uncorrected} (middle plot) shows the errors (on estimated projected iodine density) due to scattered radiation along with its dependence on sample-thickness.} The corresponding percent errors are computed with respect to the known material thickness and densities. It is apparent that the estimated iodine distributions are increasingly underestimated due to scatter as the object thickness increases. Additionally, these thickness-dependent inaccuracies are more sensitive to the higher concentration iodine targets. This occurs because the loss of primary photons due to the higher attenuating regions in the object, which in turn heightens the $SF$ within that region. Thus, the quantitative accuracy reduces as the iodine concentration is reduced as it becomes a weaker signal.
	\begin{figure}[]
		    \centering
		    \includegraphics[width=0.49\textwidth]{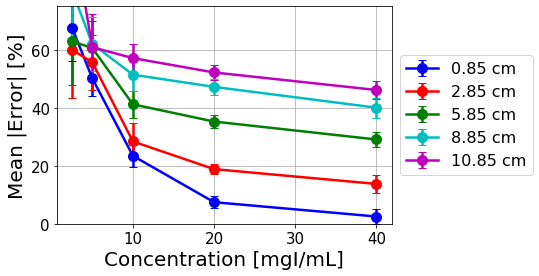}
		\caption{Mean errors of each iodine target for different object thicknesses evaluated from the image profiles in Figure \ref{fig:uncorrected}. Error bars represent the standard deviation of the percent error over each target rod.}
		\label{fig:mean_error_uncorrected}
	\end{figure}

A summary of the mean errors for each target concentration and object thickness is provided in Figure \ref{fig:mean_error_uncorrected}. The measured results are fairly accurate for thinner samples and higher iodine concentrations, observing a $\sim$3.3\% mean error for the 40 mg/ml target rod with the 0.85 cm object thickness (phantom-only). The mean errors increase nearly exponentially as the concentration is reduced, up to 21\% for the 10 mg/ml target rod and 44\% for the 5 mg/ml target rod for the same 0.85 cm object thickness. 

Figure \ref{fig:mean_error_uncorrected} also shows that the measurement accuracy diminishes as the object thickness increases. Generally, the primary intensity reduces at a faster rate than the scatter intensity as the object thickness increases, contributing to an increasing scatter fraction for thicker objects. The thickness-dependent error is more severe for the higher concentrations iodine targets, in which the increased absorption of the iodine rods further increases the scatter fraction. In the case of the 40 mg/ml target rod, the mean error increases by 25\% and 42\% when increasing the object thickness from 0.85 cm to 5.85 cm and 10.85 cm, respectively. On the other hand, the mean error of the 5 mg/ml target rod increases by only 10-12\% when increasing the object thickness from 0.85 cm to 5.85 cm and 10.85 cm. Accounting for these large variations due to relatively small highly attenuating targets may be difficult since they are generally not distinguishable from the soft tissue background. Our proposed scatter correction technique (results shown in the next section) uses the isolated soft tissue projected density to generate an approximated scatter distribution due to the soft tissue bulk material for the specified energy windows in the dual energy subtraction. These scatter distributions are then scaled by the uncorrected intensity acquisitions to accurately account for the increased scatter fractions behind small higher attenuating targets.

\subsection{Proposed scatter correction technique}
%	Target visibility and quantification may be restored by accounting for the scatter radiation in a correction technique. One requirement is estimating the energy-dependent scatter fraction which may be altered based on the energy window selection of the PCD. Additionally, the scatter correction technique has to account for the spatially-varying scatter fraction which depends on the object and target distribution.
		
\begin{figure}[]
		    \centering
		    \includegraphics[width=0.49\textwidth]{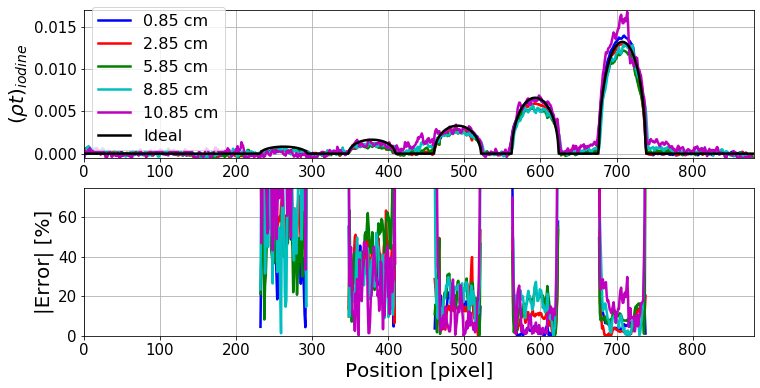}
		\caption{Similar to \textcolor{black}{the middle and bottom plots of} Figure \ref{fig:uncorrected}, but after digitally removing the scattered radiation within the total measured intensity. Quantitative accuracy of the iodine targets are restored for each sample thickness.}
		\label{fig:corrected}
	\end{figure}

\textcolor{black}{The results of applying the proposed scatter correction technique (described in Section \ref{sec:theory}) to the dual-energy subtraction is presented. In review, the spectral measurements ($\mu(E_1)$ and $\mu(E_2)$ in Figure \ref{fig:mut}) are used as inputs of Equation \ref{eq:bg_map} to get the tissue density map ($[\rho t]_b$). This result is then used in Equation \ref{eq:is} as the tissue density map (($[\rho t(\pmb{x})]^{tissue}$) for obtaining the estimated scatter distribution for each energy bin of interest (represented by $(E_1)$  and $(E_2)$). These distributions are subtracted from the corresponding measured intensities resulting in scatter corrected intensities. From these, $\mu(E_1)$ and $\mu(E_2)$ (for the known object thickness) are obtained using Beer-Lambert's law. Using these maps in Equation \ref{eq:rhoc} yields scatter-corrected density map of the contrast agent within the object. As described in the last section, prior knowledge about the iodine free regions allow a baseline subtraction to account for the non-ideal detector spectral response.} Figure \ref{fig:corrected} shows the scatter-corrected image profiles of these estimated projected iodine densities (and their corresponding percent errors with the ideal projections). The scatter-corrected target profiles are enhanced and better approximate the known ideal distributions for each rod and object thickness. Unlike the scatter-corrupted estimations (see Figure \ref{fig:uncorrected}), the thickness-dependent inaccuracies are also corrected for. 

	\begin{figure}[]
		    \centering
		    \includegraphics[width=0.49\textwidth]{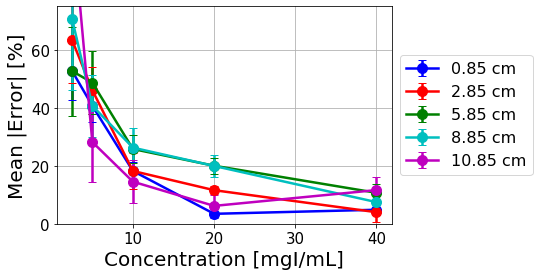}
		\caption{Mean errors of each iodine target for different object thicknesses evaluated from the scatter-corrected image profiles in Figure \ref{fig:corrected}. Error bars represent the standard deviation of the percent error over each target rod.}
		\label{fig:mean_error_corrected}
	\end{figure}	
		
Figure \ref{fig:mean_error_corrected} summarizes the mean errors for the scatter-corrected iodine distributions. The mean errors for each object thickness remain below 22\% for iodine concentrations 10 mg/ml and above. This is a considerable improvement from the uncorrected profiles, where the mean errors of the 10 mg/ml iodine target exceeds 21\% (up to 58\%) for all object thicknesses (see Figure \ref{fig:mean_error_uncorrected}). 

In addition to the improved target quantification, the scatter-corrected iodine projections exhibit far less sensitivity to the object thickness (see Figure \ref{fig:mean_error_corrected} in comparison to Figure 8). As an example, in the absence of scatter correction, the mean estimation error for 20mg/ml target rod varied by 43\% with a change in the object thickness from 0.85cm to10.85cm. Applying our scatter correction method tightened this range to 16\% while also reducing the overall estimation errors in all cases. Similar improvements were observed for other higher concentration targets, such as for 10mg/ml. In this case, as the background thickness changed from 0.85cm to10.85cm, the variations in iodine density estimation errors went from 31.5\% to 12\% respectively without and with scatter correction. As described previously, the lower concentration rods did not show an equivalent prominent thickness dependence, and showed no reduction to the thickness-dependency of the error after scatter correction.
		
\begin{figure}[]
		    \centering
		    \includegraphics[width=0.49\textwidth]{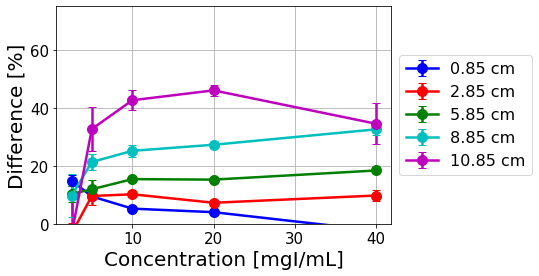}
		\caption{Percent improvement estimated by evaluating the difference in the mean errors before and after applying the proposed scatter correction technique for each target concentration and object thickness. The error bars represent the standard deviation of the mean error difference along each target rod.}
		\label{fig:percent_improvement}
	\end{figure}

Figure \ref{fig:percent_improvement} highlights the improvements to the iodine distribution estimations by evaluating the difference in the mean errors before and after scatter compensation. Improvements consistently increase as the object thickness increases, reflecting the mitigation of the scattered radiation. Error reductions of up to 46\% were observed (for the 20 mg/ml target rod within the 10.85 cm object thickness). The error reductions were also fairly consistent over the 10--40 mg/ml target rods for a given object thickness, indicating the increased scatter fractions behind higher absorbing targets was properly corrected for. 

Reduced benefits are observed as the concentration is lowered below 10 mg/ml and for sample thicknesses $\geq$2.85 cm. Recovering an accurate distribution of these low iodine concentrations may require more aggressive noise reduction to improve the estimation. In fact, improvement of the lower target concentrations is observed for the thinner 0.85 cm object which exhibits less quantum noise. We note that the 0.85 cm thickness uniquely shows a negative improvement for the 40 mg/ml target. This is disregarded as the 40 mg/ml iodine profile showed superior accuracy prior to scatter correction, and resulted in a very slight overestimation after scatter correction. 	

\begin{figure}[]
		    \centering
		    \includegraphics[width=0.49\textwidth]{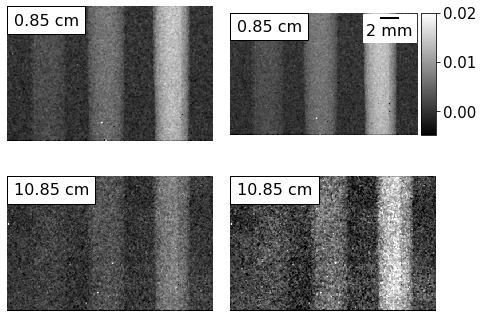}
		\caption{(left) Original and (right) scatter-corrected iodine projections of the 10, 20, and 40 mg/ml targets, from left to right, shown for varying object thickness.}
		\label{fig:quality}
		\end{figure}

\begin{figure}[]
		    \centering
		    \includegraphics[width=0.45\textwidth]{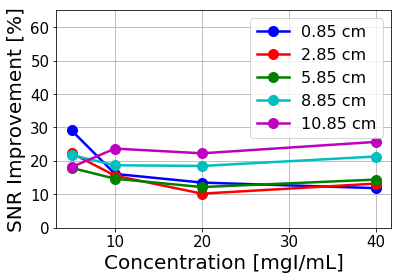}
		\caption{SNR improvement estimated by evaluating the difference before and after applying the proposed scatter correction technique for each target concentration and object thickness.}
		\label{fig:snr}
		\end{figure}

As a final evaluation of the scatter correction performance, Figure \ref{fig:quality} compares the uncorrected and corrected iodine maps. In the uncorrected case, the iodine rod signal shows reduction with increasing object thickness. This effect is reduced in the scatter-corrected case. However, there appears to be some noise amplification in the scatter-corrected case. The signal-to-noise improvement was evaluated for each image profile in Figure \ref{fig:snr}. Similar to the iodine distribution estimation accuracy, the SNR generally improves  as the object thickness is increased and is fairly consistent over the 10--40 mg/ml target rods. SNR improvements up to 27\% are observed for the 10.85 cm object thickness.

%%%%%%%%%%%%%%%%%%%%%%%%%%%%%%%%%%%%%%%%%%%%%%%%%%%%%%%%%%%%%%%%%%%%
% DISCUSSION AND CONCLUSION
%%%%%%%%%%%%%%%%%%%%%%%%%%%%%%%%%%%%%%%%%%%%%%%%%%%%%%%%%%%%%%%%%%%%%
\section{DISCUSSION}
Methods for measuring, modeling and compensating for scatter-resultant errors can prove critical in improving quantitative accuracy for applications where spectral data is captured using PCDs. We have proposed a scatter correction method that accounts for the energy-dependent characteristics of the scattered radiation. Further, we examined the influence of scatter radiation on the quantitative accuracy of iodine contrast agent mapping in a breast-like phantom. When left uncorrected, the iodine distributions are increasingly underestimated as the object thickness increases. Errors up to 28\% and 45\% are observed for the highest tested iodine concentration (40 mg/ml) in 5.85 cm and 10.85 cm object thickness, respectively. The errors increase nearly exponentially as the iodine concentration is reduced, which limits the minimum contrast agent dose delivered to a patient for adequate detectability. 

The spectral scatter estimation model described in Equation \ref{eq:is}, which has been adapted from previous literature \cite{dinten2004x, rinkel2007new} in conventional EID x-ray imaging, has reduced the quantitative inaccuracies due to the scatter.The error was reduced by up to 46\% by apply the scatter correction. The thickness dependency of the error prior to the scatter correction was greatly reduced after applying the scatter correction, ranging between 43\% before correction and 16\% after correction for the 20 mg/ml rod over the 0.85 - 10.85 cm object thickness. 

 \textcolor{black}{In this work, we focused on identifying and correcting for scatter when using PCDs. However, quantitative errors in these measurements can result from both object scatter and detector spectral distortions. As shown here, with sufficient prior information, simple methods can be used to compensate for spectral distortions, thereby isolating object scatter related errors. Methods to separate as well as compensate for both spectral and scatter errors for any general PCD based imaging applications would be part of our future research.}

Our proposed scatter correction approach is appealing since it does not require any modifications to the acquisition process or additional components to the beam-path. Rather, this method relies on physically modeling the scatter interactions and an off-line scatter database of homogeneous PMMA slabs. Unlike \cite{rinkel2007new} who prepared the database through beam-stop array, we used Monte Carlo simulation to accurately account for the spectrally-varying scatter fractions \cite{lewis2019spectral}. \textcolor{black}{Although we considered constant thickness objects (as relevant to the breast compression in clinical mammography), this method may be expanded to prepare a scatter database of non-uniform thickness objects.}
	
The low technical demand of this scatter correction technique may make it flexible for other imaging modalities. Effective image improvement has been previously shown in conventional CBCT \cite{rinkel2007new}. This model may also benefit scatter correction for digital breast tomosynthesis, where the use of an anti-scatter grid is not practical. Additionally, although this study used multiple acquisitions with a stepping energy threshold, more sophisticated acquisitions can be performed with a single spectroscopic acquisition, which reduces patient dose and minimizes motion artifacts.

\section{Conclusion}
As PCDs are being explored for medical and industrial imaging applications, there is a critical need to understand spectral characteristics of scattered x-ray photons. This paper presents a scatter correction method that accounts for the energy-dependent characteristics of the scattered radiation in spectral imaging with PCDs. We have demonstrated the scatter correction performance using a dual-energy application in contrast-enhanced mammography. The mean error of the thickest sample was reduced from $\sim$50\% to $\sim$10\% for iodine concentrations as low as 5 mg/ml.

\appendices

\section*{Appendix}\label{sec:des_derivation}

\subsection{Mathematical formalism of dual-energy subtraction:}\label{app:des}
 The energy-dependent attenuation coefficient $\mu(E)$ through an object of thickness $t$ can be determined by the Beer-Lambert law:
	\begin{equation}\label{eq:beer}
		\mu(E)t = -\ln\bigg(\frac{I_p(E)}{I_0(E)}\bigg),
	\end{equation}
	where $I_p(E)$ is the primary radiation transmitted through the object without interaction and $I_0(E)$ is the radiation distribution at the detector plane in the absence of the object. 
	
	In contrast-enhanced x-ray imaging, the total projected attenuation can be approximated as a combination of the background material ($b$) and contrast agent ($c$):
	
	\begin{equation}\label{eq:basis_approx}
	\mu(E)t = \bigg(\frac{\mu(E)}{\rho}\bigg)_b(\rho t)_b + \bigg(\frac{\mu(E)}{\rho}\bigg)_c(\rho t)_c,
	\end{equation}
	where $\rho$ is the material density. Spectroscopic imaging at two energy realizations ($E_1$ and $E_2$) yield the following system of equations \cite{baldelli2006evaluation, sarnelli2007quantitative, pani2017high}:
	\[
	\begin{bmatrix}
		\mu_1t \\
		\mu_2t
	\end{bmatrix}
	=
	\begin{bmatrix}
		(\frac{\mu_1}{\rho})_{b}	&	(\frac{\mu_1}{\rho})_{c} \\
		(\frac{\mu_2}{\rho})_{b}	&	(\frac{\mu_2}{\rho})_{c}
	\end{bmatrix}
	\begin{bmatrix}	
		(\rho t)_b \\
		(\rho t)_c	
	\end{bmatrix},
	\]
	where $\mu_i = \mu(E_i)$.
	
The density projections of the background and contrast materials can be estimated by:
	\[\begin{bmatrix}	
		(\rho t)_b \\
		(\rho t)_c	
	\end{bmatrix}
	=
	\begin{bmatrix}
		(\frac{\mu_1}{\rho})_b	&	(\frac{\mu_1}{\rho})_c \\
		(\frac{\mu_2}{\rho})_b	&	(\frac{\mu_2}{\rho})_c
	\end{bmatrix}^{-1}
	\begin{bmatrix}
		\mu_1 t \\
		\mu_2 t
	\end{bmatrix}\]
	\[
	\hspace{5em}= \frac{1}{D}
	\begin{bmatrix}
		(\frac{\mu_2}{\rho})_c	&	-(\frac{\mu_1}{\rho})_c \\
		-(\frac{\mu_2}{\rho})_b	&	(\frac{\mu_1}{\rho})_b
	\end{bmatrix}
	\begin{bmatrix}
		\mu_1 t \\
		\mu_2 t
	\end{bmatrix}, 
	\]
	where $D=(\frac{\mu_1}{\rho})_b(\frac{\mu_2}{\rho})_c - (\frac{\mu_1}{\rho})_c(\frac{\mu_2}{\rho})_b$. An alternative form of this expression can be written as: 
	\begin{eqnarray}
	(\rho t)_b &=& \frac{\big(\frac{\mu_2}{\rho}\big)_c (\mu_1t) - \big(\frac{\mu_1}{\rho}\big)_c (\mu_2t)}{D},  \label{eq:bg_map}\\
	(\rho t) _c &=& \frac{\big(\frac{\mu_1}{\rho}\big)_b (\mu_2t) - \big(\frac{\mu_2}{\rho}\big)_b (\mu_1t)}{D}, \label{eq:c_map} \label{eq:rhoc}
	\end{eqnarray}
	
\begin{figure}
		    \centering
		  	\includegraphics[width=0.45\textwidth]{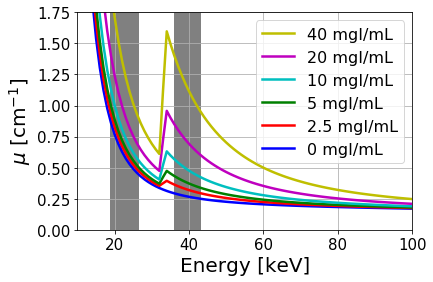}
		    \label{fig:}   
		\caption{Spectrally-varying linear attenuation coefficients for different concentrations of iodine in water. The highlighted regions correspond to the energy windows used in this study.}
		\label{fig:lac}
	\end{figure}

	The projected densities $\rho t$ can then be estimated using the mass attenuation coefficients $\frac{\mu(E)}{\rho}$ (available from the NIST XCOM database \cite{berger2010xcom}) and the measured attenuation projections $\mu(E)t$, as demonstrated in Figure \ref{fig:de_result}. This method relies on the spectral attenuation properties of background and contrast agent being distinct enough to separate the two materials. Typical contrast agent solutions contain a material with a K-edge absorption peak within the acquisition energy range, such as iodine or gadolinium. Figure \ref{fig:lac} shows the distinct linear attenuation profiles for different concentrations of iodine in water.

	Dual-energy subtraction with conventional EIDs require multiple acquisitions using two different x-ray source realizations, usually by varying the tube potential and/or filtering. An alternative solution is replacing the EID with an energy-resolving PCD which can acquire spectroscopic information about the sample using a single x-ray source realization \cite{pani2012optimization, pani2017high}.
	
\subsection{Impact of scattered radiation in DES} Ideally, the measured intensity would consist only of the primary radiation. However, radiation that scatters from the object ($I_s$) and reaches the detector obscures the primary radiation and contributes the total measured intensity. Defining $\hat{\mu}$ as the estimated linear attenuation coefficient that has been corrupted by the scattered radiation and $\mu$ as the estimated linear attenuation coefficient under ideal (scatter-free) conditions, the scatter radiation can be shown to reduce the measured attenuation coefficient by:

	\begin{equation}
	\hat{\mu} t = -\ln{\frac{I_p + I_s}{I_0}} = \mu t  + \ln(1-SF),
	\end{equation}
	where $SF=\frac{I_s}{I_p + I_s}$ is the scatter fraction. The scatter fraction is always less than one and therefore results in underestimation of the measured attenuation as $SF$ increases. 
	
Accurate material separation using dual-energy subtraction (see Equations \ref{eq:bg_map} and \ref{eq:c_map}) relies on a precise acquisition of the primary radiation to obtain the ideal $\mu$. Replacing the ideal $\mu$ with the scatter-corrupted $\hat{\mu}$ results in scatter-corrupted density projections $(\hat{\rho t})_b$ and $(\hat{\rho t})_c$ described by:

%	\begin{align}
%		(\hat{\rho t})_b &= (\rho t)_b + \frac{\mu_c(E_2) \ln(1-SF(E_1)) - \mu_c(E_1)\ln(1-SF(E_2))}{D}, \\
%		(\hat{\rho t})_c &= (\rho t)_c + \frac{\mu_b(E_1) \ln(1-SF(E_2)) - \mu_b(E_2)\ln(1-SF(E_1))}{D}.
%	\end{align}
	\begin{align}
		(\hat{\rho t})_b &= (\rho t)_b + \frac{\mu_c(E_2) \alpha(E_1) - \mu_c(E_1)\alpha(E_2)}{D}, \\
		(\hat{\rho t})_c &= (\rho t)_c + \frac{\mu_b(E_1) \alpha(E_2)) - \mu_b(E_2)\alpha(E_1)}{D},
	\end{align}
	where $\alpha(E) = \ln(1-SF(E))$. These expressions provide valuable insight of the scatter influence on the estimated density projections. It is clear that the energy-dependence of attenuation and scatter fractions will impact the inaccuracies in the estimated density projections. 
	
	A particular focus in placed on the scatter radiation impact on the estimated density projection of the contrast agent ($(\hat{\rho t})_c$), which is often the goal in clinical CESM to identify risk factor targets. Since the term containing $SF$ is always negative, $\mu_b(E_1) \ln(1-SF(E_2))$ results in an underestimation and $\mu_b(E_2)\ln(1-SF(E_1))$ results in an overestimation of the measured contrast agent density projection. The relative magnitude these competing inaccuracies will be described by considering $E_1$ to be the lower energy realization and $E_2$ to be the higher energy realization. The attenuation coefficient of the low-Z background reduces as the energy increases, meaning $\mu_b(E_1) > \mu_b(E_2)$. Similarly, our previous work \cite{lewis2019spectral} has shown that the scatter fraction reduces as the energy increases, meaning $SF(E_1) > SF(E_2)$. However, it can be determined that the underestimation due to the $\mu_b(E_1)$ and $SF(E_2)$ term has a larger impact than the overestimation due to the $\mu_b(E_2)$ and $SF(E_1)$ term. In total, this results in a net underestimation of the estimated density projection $(\hat{\rho t})_c$ for increasing scatter conditions.

\section*{Acknowledgment}
We are thankful for discussions with several Medipix collaboration (CERN, Geneva) members. The authors would like to thank Ian Harmon for his assistance in collecting some additional data for validating initial results. 

\bibliographystyle{IEEEtran}
\bibliography{library_2}

\end{document}